    \documentstyle[12pt]{report}

    \newcommand{\be}{\begin{equation}}
    \newcommand{\ee}{\end{equation}}
    \newcommand{\n}{\noindent}
    \newcommand{\vs}{\vspace{0.5cm}}
    \newcommand{\f}{\frac}
    \newcommand{\ba}{\begin{eqnarray}}
    \newcommand{\ea}{\end{eqnarray}}
    \newcommand{\R}{\mbox{I} \! \mbox{R}}

 \Large

\begin{document}
    \hsize=17truecm

    \begin{center}
    {\LARGE{\bf Decoherence in a Two-Particle Model}}

    \vspace{2cm}

    {\bf  Detlef Duerr}\\ {\em Mathematisches Institut, Universitaet
    Muenchen, Germany\\ e-m: \rm duerr@rz.mathematik.uni-muenchen.de}

    \vspace{1cm}

    {\bf Rodolfo Figari}\\ {\em Dipartimento di Scienze Fisiche,
    Universit\'a di Napoli, Italy\\ e-m: \rm figari@na.infn.it}

    \vspace{1cm}

    {\bf Alessandro Teta}\\ {\em Dipartimento di Matematica Pura e
    Applicata, Universit\'a di L'Aquila, Italy} \\ {\em e-m: \rm
    teta@univaq.it }

    \end{center}

    \vspace{2cm}

    \begin{center}
    {\bf Abstract}
    \end{center}

    \vspace{0.2cm}

    \n
    We consider a simple one dimensional quantum system consisting of
    a heavy and a light particle interacting via a point interaction.
    The initial state is chosen to be a product state, with the heavy
    particle described by a coherent superposition of two spatially
    separated wave packets with opposite momentum and the light
    particle localized in the region between the two wave packets.

    \n
    We characterize the asymptotic dynamics of the system in the limit
    of small mass ratio, with an explicit control of the error. 
    We derive the corresponding reduced density matrix for the heavy
    particle and explicitly compute  the (partial)
    decoherence effect for the heavy particle induced by the presence
    of the light one for a particular set up of the parameters.

    \newpage

    \setcounter{chapter}{1} \setcounter{equation}{0}

    {\bf 1. Introduction}

    \vs

    \n
    Decoherence has become the terminology for the irreversible
    suppression of interference of the wavefunction of a quantum
    system due to the interaction with an
    "environment" ([GJKKSZ], [BGJKS]).

    \n
    The usual picture of
    decoherence, in the simple setting of a two particle system,  goes as
    follows. Suppose  one has a particle $M$ with
    initial wave function $\varphi(x)= \varphi_l(x) + \varphi_r(x)$,
    representing the superposition of two wave packets $\varphi_l$,
    more or less supported "on the left of the origin" and heading to the
    right, and
    $\varphi_r$ supported more or less on the "right of the origin"
    with an average velocity pointing to the left. Suppose that
    another particle $m$, described initially by the wave packet
    $\Phi(y)$, passes by and interacts with $M$.  Assuming a small mass
    ratio between the second and the first particle, it is conceivable
    that the evolution of $M$ will not be much affected by the
    interaction, while the scattering process undergone by the particle
    $m$ will depend strongly on the position of the heavier particle.
    After interaction (which is assumed to be very fast) one then
    expects that the wave function describing the state of the system
    is of the type $\psi(x,y)= \varphi_l(x)
    \Phi_l(y)+\varphi_r(x)\Phi_r(y)$, where $\Phi_l$ and $\Phi_r$ will
    have spatial supports concentrated in distant regions for all later times.
    Therefore in the configuration space of the entire system the
    entangled state will appear as the sum of two disjoint components and
    the possibility of interference of the heavy particle wave packets
    will be reduced.

    \n
    Notice that the reduced
    density matrix of the particle $M$ has in this ideal case
    negligible off-diagonal elements. In this sense, interference has
    been reduced and the motion of the particle $M$ has become more
    "classical". That is the way decoherence plays a role in the
    explanation of the emergence of classical behaviour from quantum
    mechanics. 

\n
For the relevance of the mechanism of decoherence in the classical limit
of Quantum Mechanics in the language of Bohmian Mechanics see [ADGZ].

    \n
    In this respect it is an interesting  problem to separate "pure
    decoherence"  from the other effects which an environment usually
    produces, which are dissipation and fluctuation. That is, one
    would like to have the motion of the system not much affected by
    the interaction with the environment, while the environment
    produces decoherence. It is unclear whether these desiderata,
    namely a good decoherence rate and a more or less unperturbed
    motion can be consistently fullfilled in realistic physical models
     ([GH]).

    \n
    Explicit models where one can rigorously
    establish decoherence in this sense  have been worked out in the
    last years. One such model has been studied in ([DS]) where
    the interaction of a particle with the radiation field has been
    considered. We shall now study another interaction which
    elaborates closer the idea of scattering of light particles (the
    environment) off a heavy particle (the system) (see
     [JZ],[GF],[T] for similar ideas).

    \n
    We consider a very simple one dimensional model of a system (a
    heavy particle of mass $M$) plus environment (one light particle
    of mass $m$ ) interacting via a short range force
    ($\delta-$interaction). We consider this case as  useful
    preparation for the treatment of a three dimensional gas of light
    particles interacting with the heavy particle, which we shall
    address in subsequent work.

    \n
    We wish to stress some features that make the two-particle
    model with  $\delta-$interaction (which has the advantage of
    being analytically easily accessible) particularly suitable as
    a model for decoherence.

    \n
    \begin{enumerate}
    \item
    there exists a simple dimensionless parameter in the problem,
    namely the fraction of the masses $\epsilon=\frac{m}{M}$.
    \item
    Letting $\epsilon$ become small while keeping $M$ fixed allows to
    approximate the solution  of the Schr\"odinger equation of the two
    body problem  by a scattering solution in which the heavy particle
    acts as a scattering center for the light one. The error is ${\cal
    O}(\epsilon)$. The time scale on which this approximation holds is
    of course given by the time the light particle needs to pass the
    heavy particle.  This approximation is the starting point of the
    analysis in [JZ].

    \item
    The decoherence effect (i.e. the amount by which the off diagonal elements
    of the reduced density matrix are reduced) can be explicitely
    computed (see (\ref{decoheff1}) and the discussion follwing it)
    and is, in the relevant regime, of the order of $\alpha_0
    m \hbar^{-2} \delta$, where $\delta$ is the initial spread of the light
    particle and $\alpha_0$ is the strength of the potential
    ($(\alpha_0 m)^{-1} \hbar^2$ is the effective range of interaction) all of
    which can also be chosen $\epsilon$-dependent.

    \end{enumerate}

    \n
    We wish to warn
    the reader, that the point interaction we look at here in form of
    the $\delta$-potential is for finite $\alpha_0$ not a hard core
    interaction. The case $\alpha_0\to \infty$ corresponds to hard
    core.

    The paper is organised as follows.

    \n
    In Section 2 we introduce the model and characterise the
    asymptotic dynamics of the two-particle system for small mass
    ratio and state the main approximation result.

    \n
    In Section 3 we show the attenuation of the
    off-diagonal terms in the reduced density matrix for the heavy
    particle and we compute explicitly the probability distribution
    for the position of the heavy particle, showing reduction of the
    interference effects with respect to the non interacting case.

    \n
    In Section 4 we give the proof of the main result of the paper.

    \n
    In the appendix we recall the derivation of the explicit solution
    of the Schroedinger equation of the two-body system in interaction
    via a delta potential in dimension one.

    \vspace{1cm} \setcounter{chapter}{2} \setcounter{equation}{0}

    {\bf 2. Expression for small mass ratio}

    \vs
    \n
    In this section we shall study   the Schroedinger equation for the
    two-particle system in one dimension described by the hamiltonian

    \ba &&H= - \frac{\hbar^{2}}{2M} \Delta_{R} - \f{\hbar^{2}}{2m}
    \Delta_{r} + \alpha_0 \delta (r-R), \hspace{2cm} \alpha_0 >0
    \label{ham} \ea

    \n
    In (\ref{ham}) we have denoted by $R$ the position coordinate of
    the heavy particle with mass $M$ and by $r$ the position
    coordinate of the light particle with mass $m$. The interaction
    potential is chosen to be a repulsive point interaction of strenth
    $\alpha_0$.

    \n
    It is well known that (\ref{ham}) is a well defined positive and
    selfadjoint operator in $L^{2}(\R^{2}, drdR)$, which is also a
    solvable model ([AGH-KH]).

    \n
    In fact, for an arbitrary initial state $\psi_{0}=\psi_{0}(r,R)$,
    the solution of the  Schroedinger equation can be explicitely
    written as  (see [S] and the appendix)

    \ba &&\psi(t,r,R)= \int dr' dR' \psi_0(r',R') U_{0}^{\nu} \left(t,
    \f{M}{\nu}(R-R')+ \f{\mu}{M}(r-r')\right) \nonumber\\
    &&\nonumber\\ && \cdot \left[   U_{0}^{\mu}(t,(r-R)-(r'-R'))
     - \f{\mu \alpha_0}{\hbar^{2}}
    \int_{0}^{\infty} du e^{- \f{\mu \alpha_0}{\hbar^{2}} u}
    U_{0}^{\mu}(t,u+|r-R|+|r'-R'|) \right] \nonumber\\ &&
    \label{sol.esatta} \ea

    \n
    where we have introduced the reduced mass and the total mass of
    the system

    \be
     \mu=\f{mM}{m+M}, \;\;\;
    \nu=m+M \ee

    \n
    and the integral kernel of the free unitary group $U_{0}^{\cal
    M}(t)$ corresponding to the mass ${\cal M}>0$

    \be
    U_{0}^{\cal M}(t,x-x') = e^{-i \f{t}{\hbar} H_{0}^{\cal M}}(x-x')
    = \sqrt{\f{\cal M}{2 \pi i \hbar t}} e^{i \f{\cal M}{2 \hbar  t}
    (x-x')^{2}}, \;\;\;\; x,x'\in \R \ee

    \n
    We are interested in the case of an initial state in a product
    form. Then we fix two real valued smooth functions (for ease of formulation 
we
  assume that they are in Schwartz space $\cal{S}$)

   \be
    f,g \in {\cal{S}}, \;\; \;\;\;\; \|f\|=\|g\|=1 \ee

    \n
    where $\| \cdot\|$ denotes the norm in $L^{2}(\R)$. For later use,
    it will be convenient to choose $g$ even. Using $f$ and $g$ we
    define now the states in such a way that we can easily read of the
    relevant physical scales, i.e. we code the states by the physical
    parameters $R_0, P_0,   \sigma, r_0, q_0, \delta$ as follows

    \ba &&f_{\sigma,  R_{0},  P_{0}}(R) = \f{1}{\sqrt{2}} \left[
    f_{\sigma, R_{0},  P_{0}}^{+}(R) + f_{\sigma,  R_{0},
    P_{0}}^{-}(R) \right] \label{f}
    \\
    &&f^{\pm}_{\sigma, R_{0}, P_{0}}(R)= \f{1}{\sqrt{\sigma}} f \left(
    \f{R \pm R_{0}}{\sigma} \right) e^{\pm i \f{P_{0}}{\hbar} R}
    \label{fpm}
    \\
    &&g_{\delta, r_0, q_0} (r) = \f{1}{\sqrt{\delta}} g \left(
    \f{r-r_0}{\delta} \right) e^{i \f{q_0}{\hbar} r} \label{g}\\
    &&\sigma, \delta, R_0, P_{0}, q_0 >0, \;\; r_0 \in \R, \;\;\;\; R_0 >
    \sigma + \delta +
    |r_0|\label{param}
    \ea

    \n
    The choice (\ref{param}) is not essential for the most part of the
    paper, but it sets already a geometrical picture which puts the
    results in the right perspective (see below (\ref{statoin})).
    Later on we shall use this particular choice for computing
    effects. Note that the spread of the wave function of $M$ is not
    given by $\sigma$ but by $R_0$.

    \n
    The initial state that we consider in the following is

    \be
    \psi_0 (r,R) =  g_{\delta, r_0, q_0} (r) f_{\sigma,  R_{0},
    P_{0}}(R) \label{statoin} \ee

    \n
    The initial state (\ref{statoin}) is a (pure) product state for
    the whole system, i.e. no correlation is assumed between the two
    particles at time zero.

    \n
    The heavy particle is assumed to be in a  superposition  of two
    spatially separated  wave packets, one localized in $R=-R_{0}$
    with mean value of the momentum $P_{0}$ and the other localized in
    $R=R_{0}$ with mean value of the momentum $-P_{0}$. The light
    particle is localized around $r_0$, in the region between the two wave
    packets,   with positive mean
    momentum $q_0$.

    \n
    To simplify the notation, in the rest of the paper we shall drop
    the dependence of the initial state on  $R_{0},  P_{0}, r_0, q_0$.
    Moreover, for the convenience of the reader, we collect here some
    notation which will be used later on

    \ba
    &&\epsilon  = \f{m}{M},\;\;\;\; \mu = \f{\epsilon}{1+\epsilon}
    M,\;\;\;\; \nu = (1+\epsilon) M\label{epsilon}\\ &&\alpha  =
    \f{\alpha_{0} m }{\hbar^{2}}
    \label{alpha0}\\ &&k_0 = \frac{q_0}{\hbar}, \;\;\;\;\;\; K=
    \f{P_0}{\hbar} + k_0\label{k0}\\ &&T\; : \; L^{2}(\R^{2},drdR)
    \rightarrow L^{2}(\R^{2}, dx_{1}dx_2), \nonumber\\ &&(T h
    )(x_1,x_2) \equiv h \left(x_2+ \f{M}{m+M} x_1, x_2 - \f{m}{m+M}
    x_1 \right) \label{T}\\ &&\Delta^{\pm} = (\pm R_{0} - \sigma, \pm
    R_{0} + \sigma) \label{Deltapm} \ea

    \n
    and finally $c$ will denote a positive numerical constant.

    \n
    We shall now characterize  the asymptotic behaviour of the wave
    function for small value of the mass ratio $\epsilon$ for the
    initial state (\ref{statoin}). Letting $m$ become small, keeping
    $M$ fixed, the light particles spreads with speeds $v\sim
    \hbar/\delta m$ and the time by which the light particle passes
    $M$ is of the order of $R_0/v$ thus decreases with $m$, so that
    $M$ does not change much its position during the passing of $m$.

    \n
    The limit dynamics will hence describe a situation in which the
    light particle is scattered  by the heavy one being in some fixed
    position, while the heavy particle moves freely.
    \n
    Nevertheless, we shall find that the  free motion of the heavy
    particle is modified by the scattering event. In this heuristic
    argument we kept all the other physical parameters fixed except
    for the interaction strength. In fact, in order to keep the
    interaction effective on the light particle we need to scale $\alpha_0$
    in such a way that $\alpha_0 m \approx {\cal O}(1)$. There is of course no
    need to keep the other parameters fixed, in fact one may well
    imagine $\delta$ and $R_0$ increasing with $m$, so that the
    kinetic energy of $m$ stays finite and the spread of the $M$
    increases. We shall not discuss such choices here, but the
    estimates are detailed enough, so that other scalings can be
    easily discussed. This might become relevant in a model where the
    heavy particle is immersed in a gas of light particles.

    \n
    In order to formulate the main result of this section, we define
    the integral operator

    \ba &&(W_{+}^{\gamma,x_0}h)(k) = \f{1}{\sqrt{2\pi}} \int dx h(x)
    \left( e^{-i kx} +{\cal R}_{\gamma}(k) e^{-i x_0 k} e^{i |k| |x-
    x_0|} \right), \;\;\;\;\; \gamma>0,\; x_0 \in \R \;\;\;\;
    \label{W} \\
    && {\cal R}_{\gamma}(k) =- \f{\gamma}{\gamma - i |k|}
    \label{cal R} \ea

    \n
    where the integral kernel in (\ref{W}) is the generalized
    eigenfunction of the hamiltonian

    \be
    H_{\gamma,x_0}= - \f{1}{2} \Delta + \gamma \delta(\cdot -x_0)
    \label{H_gamma} \ee

    \n
    and ${\cal R}_{\gamma}(k)$ is the corresponding reflection
    coefficient (see e.g. [AGH-KH]). Moreover we introduce the wave
    operator $\Omega_{+}^{\gamma,x_0}$ associated to $H_{\gamma,x_0}$,
    explicitely given by

    \be
    (\Omega_{+}^{\gamma,x_0}h)(x) = \left[ (W_{+}^{\gamma,x_0})^{-1}
    \tilde{h} \right](x) \ee

    \n
    where $\tilde{h}$ denotes the Fourier transform of $h$.

    \n
    With the above notation the asymptotic wave function, which will
    be denoted by $\psi^{a}(t)$, is explicitely characterized in  the
    following theorem.

    \vs
    \n
    {\bf Theorem 1.} {\em Given the initial state (\ref{statoin}),
    then for any $t>0$ the following estimate holds

    \be
    \| \psi (t) - \psi^{a}(t) \| < \left(\frac{A}{t} + B \right) \, \epsilon \ee

    \n
    where

    \ba &&\psi^{a}(t,r,R)= \sqrt{\f{m}{i \hbar t}} e^{i \f{m}{2 \hbar
    t} r^{2}} \int d y  f_{\sigma} (y) U_{0}^{M}(t,R-y) \left(
    W_{+}^{\alpha,y} g_{\delta} \right) \left(\f{mr}{\hbar t}
    \right)\nonumber\\ &&= \sqrt{\f{m}{i \hbar t}} e^{i \f{m}{2 \hbar
    t} r^{2}} \int d y  f_{\sigma} (y) U_{0}^{M}(t,R-y) \left[
    (\Omega_{+}^{\alpha,y})^{-1} g_{\delta} \right]^{\tilde{}}
    \left(\f{mr}{\hbar t}\right) \label{psiacompatta} \ea

    \n
    and $A$,$B$ are positive, time-independent constants whose detailed
    dependence on the physical  parameters characterising the interaction and
    the initial state will  be given in section 4.}

    \vs

    \n
    {\bf Remark 1.} Note that $\psi^{a}(t,r,R)$ is close to what we
    described in the introduction. Think of $f_{\sigma}$ as we do as
    consisting of two well concentrated  wavepackets, then we have the
    light particle's scattered wavefunction correlated with the two
    mean positions of the heavy particles, i.e. read in $ \left[
    (\Omega_{+}^{\alpha,y})^{-1} g_{\delta} \right]^{\tilde{}} $ the
    $y$ morally as the scattering center.

    \vs
    \n
    The result of theorem 1 can be  rephrased in terms of reduced
    density matrix for the heavy particle, which is defined by the
    integral operator $\hat{\rho}(t)$ in $L^{2}(\R)$ given by the
    kernel

    \ba &&\hat{\rho}(t,R,R') = \int dr \psi(t,r,R)
    \overline{\psi}(t,r,R') \ea

    \n
    We also introduce the integral operator $\hat{\rho}^{a}(t)$
    defined  by

    \ba &&\hat{\rho}^{a}(t,R,R') = \int dr \psi^{a}(t,r,R)
    \overline{\psi^{a}}(t,r,R')\nonumber\\ &&= \int dy f_{\sigma}(y)
    U_{0}^{M}(t,R-y) \int dz \overline{f_{\sigma}}(z)
    \overline{U_{0}^{M}}(t,R'-z) {\cal I}(y,z)\label{rhoa} \ea

    \n
    where

    \ba &&{\cal I}(y,z) \equiv \int dk (W_{+}^{\alpha,
    y}g_{\delta})(k) \overline{(W_{+}^{\alpha, z}g_{\delta})}(k) =
    \left( (\Omega_{+}^{\alpha,z})^{-1} g_{\delta},
    (\Omega_{+}^{\alpha,y})^{-1} g_{\delta} \right) \label{cal I} \ea

    \n
    Formula (\ref{cal I}), obtained through heuristic considerations, has been
    the main ingredient in the description of scattering induced
    decoherence in [JZ].

    \n
    Observe that, from (\ref{rhoa}),(\ref{cal I}) one has

    \ba &&\hat{\rho}^{a}(t)= U_{0}^{M}(t) \hat{\rho}^{a}_{0}
    U^{M}_{0}(-t) \ea

    \n
    where $\hat{\rho}^{a}_{0}$ is defined by the integral kernel

    \ba &&\hat{\rho}^{a}_{0}(y,z) = f_{\sigma}(y)
    \overline{f_{\sigma}}(z) {\cal I}(y,z) \ea

    \n
    It is easily seen that ${\cal I}(y,z) = \overline{{\cal I}}(z,y)$,
    $|{\cal I}(y,z)| \leq 1$ and the equality holds only if $y=z$.

    \n
    Then $\hat{\rho}^{a}_{0}$ is a self-adjoint and trace-class
    operator, with $Tr (\hat{\rho}^{a}_{0}) =1$; it is also positive
    since

    \ba &&(h, \hat{\rho}^{a}_{0} h) = \int dy \overline{h}(y) \int dz
    h(z) f_{\sigma}(y) \overline{f_{\sigma}}(z) \int dk
    (W^{\alpha,y}g_{\delta})(k) \overline{(W^{\alpha,z}
    g_{\delta})}(k)\nonumber\\ &&= \int dk \left| \int dy
    \overline{h}(y) f_{\sigma}(y) ( W^{\alpha,y} g_{\delta})(k)
    \right|^{2} \ea

    \n
    Moreover we have

    \ba &&Tr( (\hat{\rho}^{a}_{0} )^{2}) = \int dy dz
    |f_{\sigma}(y)|^{2} |f_{\sigma}(z)|^{2} |{\cal I}(y,z)|^{2} <1 \ea

    \n
    We conclude that $\hat{\rho}^{a}_{0}$ and its free evolution
    $\hat{\rho}^{a}(t)$  are density matrices describing  mixture
    states and  by Theorem 1, for any $t>0$, one has

    \ba && Tr (| \hat{\rho}(t) - \hat{\rho}^{a}(t) |) < \left(\frac{A}{t} + B
  \right)
    \,
    \epsilon \ea

    \n
    This means that in our asymptotic regime the motion of the heavy
    particle is a free evolution.

    \n
    On the other hand the presence of the light particle has a
    relevant effect, since it  produces a transition of the initial
    state of the heavy  particle from
    $\hat{\rho}_{0}(y,z)=f_{\sigma}(y)\overline{f_{\sigma}}(z)$ to
    $\hat{\rho}^{a}_{0}(y,z)$.

    \n
    We shall see in the next section that this is the origin of the
    decoherence effect on the heavy particle.

    \n
    Finally, it is worth to mention that the dynamics of the system
    can be equivalently described by the Wigner function. From
    (\ref{rhoa}),(\ref{cal I}) we see that the asymptotic form of the
    reduced Wigner function describing the motion of the heavy
    particle is the free evolution of

    \ba &&\hat{W}^{a}_{0}(R,P) = \f{1}{2 \pi} \int dx e^{i Px}
    f_{\sigma}(R-\f{\hbar}{2}x) \overline{f_{\sigma}}(R+\f{\hbar}{2}x)
    {\cal I}(R-\f{\hbar}{2}x, R+\f{\hbar}{2}x) \ea

    \vspace{1 cm}

    \setcounter{chapter}{3} \setcounter{equation}{0} \vs

    {\bf 3. Size of Decoherence}

    \vs

    \n
    Here we discuss an application of formula (\ref{rhoa}),(\ref{cal
    I}) to a concrete example of quantum evolution and we give an
    explicit computation of the decoherence effect.

    \n
    We shall consider the  initial state (\ref{statoin}) with the
    further assumptions (which are only done for ease of presentation)

   \be
    f,g \in C_0^\infty (-1,+1), \ee
  
\n       
and

    \ba &&\sigma  \;\ll \;\f{1}{\alpha} \;\ll \; R_0 - |r_0|\, ,
    \;\;\;\;\;\;\delta \; \ll \; R_0 - |r_0|
    \label{assump} \ea

    \n
    i.e. the spreading in position of the wave packets (which are
    superposed in a gross superposition) of the heavy particle is much
    smaller than the effective range of the interaction and this, in
    turn, is much smaller than the separation between the two
    particles. Moreover the light particle is well separated from each
    wave packet of the heavy one.

    \n
    Notice that (\ref{assump}) obviously implies $\f{\sigma}{R_{0}} \ll
    1$.

    \n
    Using assumptions (\ref{assump}) we can give an estimate of the
    basic object ${\cal I}(y,z)$ for $y ,z \in \Delta^{\pm}$   and
    then we can find a more suitable expression for the reduced
    density matrix of the heavy particle.


    \n
    In order to formulate the result, we define the parameter

    \be
    \Lambda = \int dk |\tilde{g}_{\delta} (k)|^{2}  {\cal T}_{\alpha}
    (k) = 1 + \int dk |\tilde{g}_{\delta}(k)|^{2} {\cal R}_{\alpha}
    (k) \label{Lambda} \ee

    \n
    where

    \be
    {\cal T}_{\gamma}(k) = - \frac{i k}{\gamma - i k } = 1 + {\cal
    R}_{\gamma} (k), \;\;\;\;\; \gamma >0 \label{calT} \ee

    \n
    is the transmission coefficient associated to a point interaction
    of strength $\gamma$ (see e.g. [AGH-KH]). Then we have

    \vs
    \n
    {\bf Proposition 2.} {\em Assume (\ref{assump}). Then

    \ba &&\sup_{y,z \in \Delta^{\pm}} \left| {\cal I}(y,z) -1 \right|
    < c \left(  \alpha \sigma + \f{1}{\alpha (R_0 - |r_0|)} +
    \f{\delta}{R_{0} - |r_{0}|} \right) \label{stimadia}\\ &&\sup_{y
    \in \Delta^{+}, z \in \Delta^{-}} \left| {\cal I}(y,z) - \Lambda
    \right| = \sup_{y \in \Delta^{-}, z \in \Delta^{+}} \left| {\cal
    I}(y,z) - \overline{\Lambda} \right| < c  \left( \f{1}{\alpha (R_0
    - |r_0|)}+\f{\delta}{R_{0}-|r_{0}|} \right)\;\;
    \label{stimanondia} \ea



    \n
    }


    \vs

    \n
    {\bf Proof}. Using the shorthand notation $\beta = \alpha \delta$,
    we note that for $y \in \Delta^{-}$

    \ba &&\f{1}{\sqrt{\delta}}\left( W_{+}^{\alpha,y}g_{\delta}
    \right) \left(\f{k}{\delta}\right) = e^{i (k_0 \delta -k)
    \f{r_0}{\delta}} \tilde{g}(k- k_0 \delta) + {\cal R}_{\beta}(k)
    e^{i( k_0 \delta +|k|) \f{r_0}{\delta} -i (k+|k|)\f{y}{\delta}}
    \tilde{g}(|k| +k_0 \delta)\nonumber\\ && \ea

    \n
    and for $y \in \Delta^{+}$

    \ba &&\f{1}{\sqrt{\delta}}\left( W_{+}^{\alpha,y}g_{\delta}
    \right) \left(\f{k}{\delta}\right) = e^{i (k_0 \delta -k)
    \f{r_0}{\delta}} \tilde{g}(k- k_0 \delta) + {\cal R}_{\beta}(k)
    e^{i( k_0 \delta-|k|) \f{r_0}{\delta} -i (k-|k|)\f{y}{\delta}}
    \tilde{g}(|k|-k_0 \delta)\nonumber\\ && \ea

    \n
    where we have used the fact that

    \be
    \tilde{g}_{\delta}(k) = \sqrt{\delta} \tilde{g}(k \delta - k_{0}
    \delta) e^{-i (k-k_{0})r_{0}} \ee

    \n
    Then for $y,z \in \Delta^{-}$ we have

    \ba &&{\cal I}(y,z)= 1+ \int dk  |\tilde{g} (|k|+k_0 \delta)|^{2}
    |{\cal R}_{\beta}(k)|^{2}  e^{-i(k+|k|)\f{y-z}{\delta}}
    \nonumber\\ &&+\int dk  \overline{\tilde{g}}(k- k_0 \delta)
    \tilde{g}(|k|+k_0\delta) {\cal R}_{\beta}(k) e^{i
    (k+|k|)\f{r_{0}-y}{\delta}} \nonumber\\ &&+ \int dk
    \overline{\tilde{g}}(|k|+k_0\delta) \tilde{g}(k- k_0\delta)
    \overline{{\cal R}_{\beta}}(k) e^{-i
    (k+|k|)\f{r_{0}-z}{\delta}}\nonumber\\ &&=1+ \int_{0}^{\infty} dk
    |\tilde{g}(k+k_0\delta)|^{2} |{\cal R}_{\beta}(k)|^{2} \left( e^{-
    2 i k \f{y-z}{\delta}} -1 \right) \nonumber\\ &&+
    \int_{0}^{\infty} dk \overline{\tilde{g}}(k-k_0\delta)
    \tilde{g}(k+k_0\delta)  {\cal R}_{\beta}(k) e^{2ik \f{r_0 -
    y}{\delta}} \nonumber\\ &&+ \int_{0}^{\infty} dk
    \overline{\tilde{g}}(k+k_0\delta) \tilde{g}(k-k_0\delta)
    \overline{{\cal R}_{\beta}}(k) e^{-2i k \f{r_0 -
    z}{\delta}}\nonumber\\ &&\equiv 1 + a_1 +a_2 +a_3 \ea

    \n
    where we have used the identity ${\cal R}_{\beta} +
    \overline{{\cal R}_{\beta}} + 2 |{\cal R}_{\beta}|^{2}=0$ and the
    fact that $\tilde{g}$ is even.

    \n
    Using (\ref{assump}) we easily  estimate  $a_{1}$

    \ba &&|a_{1}| \leq 2 \f{|y-z|}{\delta} \int_{0}^{\infty} dk
    |\tilde{g}(k+k_{0}\delta)|^{2} k |{\cal R}_{\beta}(k)|^{2} \leq 4
    \f{\sigma}{\delta}  \int_{0}^{\infty} dk
    |\tilde{g}(k+k_{0}\delta)|^{2} k |{\cal
    R}_{\beta}(k)|^{2}\nonumber\\ &&\leq 2 \; \alpha \sigma \ea

    \n
    For the estimate of $a_{2}$ it is convenient to integrate by parts

    \ba &&|a_{2}| = \left| \f{1}{2i}  \f{\delta}{r_{0} - y}
    \int_{0}^{\infty} dk \overline{\tilde{g}}(k-k_{0}\delta)
    \tilde{g}(k+k_{0}\delta) {\cal  R}_{\beta}(k) \f{d}{dk} e^{2ik
    \f{r_{0}-y}{\delta}} \right|\nonumber\\ &&=\f{\delta}{2|r_{0} -
    y|} \left| \int_{0}^{\infty} dk \f{d}{dk} \left(
    \overline{\tilde{g}}(k-k_{0}\delta) \tilde{g}(k+k_{0}\delta) {\cal
    R}_{\beta}(k) \right) e^{2ik \f{r_{0} - y}{\delta}} -
    |\tilde{g}(k_{0}\delta)|^{2} \right|\nonumber\\ &&\leq
    \f{\delta}{R_{0}-|r_{0}|} \left[ \f{1}{2\pi} \left( \int dr |g(r)|
    \right)^{2} + \f{1}{\beta} \int_{0}^{\infty} dk \left|
    \overline{\tilde{g}}(k-k_{0}\delta)
    \tilde{g}(k+k_{0}\delta)\right| \right. \nonumber\\ &&+
    \int_{0}^{\infty} dk \left| \overline{\tilde{g}'}(k-k_{0}\delta)
    \tilde{g}(k+k_{0}\delta)\right| + \left.\int_{0}^{\infty} dk
    \left| \overline{\tilde{g}}(k-k_{0}\delta)
    \tilde{g}'(k+k_{0}\delta)\right|\right]\nonumber\\ &&\leq
    \f{\delta}{R_{0}-|r_{0}|} \left[ \f{1}{2\pi} \left( \int dr |g(r)|
    \right)^{2} + \f{1}{\beta} \|g\|^{2} + 2 \|\tilde{g}'\|
    \right]\nonumber\\ && \leq \f{\delta}{R_0 - |r_0|}
    \left(\f{1}{\pi} + 2 \|\tilde{g}'\|\right) + \f{1}{\alpha (R_0
    -|r_0|)} \label{a_{2}} \ea

    \n
    The term $a_{3}$ is analysed exacly in the same way and then we
    get the estimate (\ref{stimadia}) for $y,z \in \Delta^{-}$. Since
    in the case $y,z \in \Delta^{+}$ the computation is similar  we
    conclude that (\ref{stimadia}) holds.

    \n
    In order to prove (\ref{stimanondia}) we consider the case $y \in
    \Delta^{+}$ and $z \in \Delta^{-}$ (the case $y \in \Delta^{-}$
    and $z \in \Delta^{+}$ can be treated exactly in the same way) and
    we obtain

    \ba &&{\cal I}(y,z) = 1 + \int dk \overline{\tilde{g}}(|k| +
    k_{0}\delta) \tilde{g}(|k|-k_{0}\delta) |{\cal R}_{\beta}(k)|^{2}
    e^{-2i |k| \frac{r_{0}}{\delta} +i (|k| -k) \f{y}{\delta} + i
    (|k|+k) \f{z}{\delta}} \nonumber\\ &&+ \int dk
    \overline{\tilde{g}}(k - k_{0}\delta) \tilde{g}(|k|-k_{0}\delta)
    {\cal R}_{\beta}(k)   e^{-i (|k|-k) \f{r_{0} -  y}{\delta}}
    \nonumber\\ &&+ \int dk \overline{\tilde{g}}(|k| + k_{0}\delta)
    \tilde{g}(k -k_{0}\delta)  \overline{{\cal R}_{\beta}}(k) e^{-i
    (|k| + k) \f{r_{0} -z}{\delta}} \nonumber\\ &&= 1 +
    \int_{0}^{\infty} dk \left( |\tilde{g} (k-k_{0}\delta)|^{2} {\cal
    R}_{\beta}(k) + |\tilde{g}(k+k_{0}\delta)|^{2} \overline{{\cal
    R}_{\beta}}(k) \right)\nonumber\\ &&+ \int_{0}^{\infty} dk
    \overline{\tilde{g}}(k+k_{0}\delta) \tilde{g}(k-k_{0}\delta)
    |{\cal R}_{\beta}(k)|^{2} e^{-2ik \f{r_{0}-z}{\delta}} \nonumber\\
    &&+ \int_{0}^{\infty} dk \overline{\tilde{g}}(k+k_{0}\delta)
    \tilde{g}(k-k_{0}\delta) \overline{{\cal R}_{\beta}}(k) e^{-2ik
    \f{r_{0}-z}{\delta}}\nonumber\\ &&+ \int_{0}^{\infty} dk
    \overline{\tilde{g}}(k+k_{0}\delta) \tilde{g}(k-k_{0}\delta)
    |{\cal R}_{\beta}(k)|^{2} e^{-2ik \f{r_{0}-y}{\delta}} \nonumber\\
    &&+ \int_{0}^{\infty} dk \overline{\tilde{g}}(k+k_{0}\delta)
    \tilde{g}(k-k_{0}\delta) {\cal R}_{\beta}(k) e^{-2ik
    \f{r_{0}-y}{\delta}} \label{y+z-} \ea

    \n
    The estimate of the  last four terms of (\ref{y+z-}) proceeds
    exactly as the estimate of $a_{2}$ in (\ref{a_{2}}). On the other
    hand

    \ba && 1 + \int_{0}^{\infty} dk \left( |\tilde{g}
    (k-k_{0}\delta)|^{2} {\cal R}_{\beta}(k) +
    |\tilde{g}(k+k_{0}\delta)|^{2} \overline{{\cal R}_{\beta}}(k)
    \right)\nonumber\\ &&= \int dk |\tilde{g}(k-k_{0}\delta)|^{2}
    \left( \f{- i k}{\beta - i  k} \right) \ea

    \n
    and this concludes the proof of the proposition. $\Box$

    \vs
    \n
    Proposition 2 allows us to find a further approximate form for the
    reduced density matrix.

    \vs
    \n
    {\bf Corollary 3.} {\em Under the assumptions (\ref{assump})  and
    for any $t \geq 0$ we have

    \ba &&\left[ Tr \left( \left( \hat{\rho}^{a}(t) -
    \hat{\rho}^{f}(t) \right)^{2} \right) \right]^{1/2} < c \left(
    \alpha \sigma + \f{1}{\alpha (R_0 - |r_0|)} + \f{\delta}{R_{0} -
    |r_{0}|} \right) \ea

    \n
     where

    \ba &&\hat{\rho}^{f}(t)= U_{0}^{M}(t) \hat{\rho}^{f}_{0}
    U_{0}^{M}(-t)\\ \label{rhoft} &&\hat{\rho}^{f}_{0}(y,z)= \f{1}{2}
    f^{+}_{\sigma}(y) \overline{f^{+}_{\sigma}}(z) + \f{1}{2}
    f^{-}_{\sigma}(y) \overline{f^{-}_{\sigma}}(z) + \f{\Lambda}{2}
    f^{+}_{\sigma}(y) \overline{f^{-}_{\sigma}}(z) +
    \f{\overline{\Lambda}}{2} f^{-}_{\sigma}(y)
    \overline{f^{+}_{\sigma}}(z) \label{rhof0} \ea

    \n
    }

    \vs
    \n
    {\bf Proof.}

    \ba &&Tr \left(  \left( \hat{\rho}^{a}(t) - \hat{\rho}^{f}(t)
    \right)^{2} \right) = Tr \left( \left( \hat{\rho}^{a}_{0} -
    \hat{\rho}^{f}_{0} \right)^{2} \right)\nonumber\\ &&= \f{1}{4}
    \int dydz \left|
     f^{+}_{\sigma}(y) \overline{f^{+}_{\sigma}}(z) ({\cal I}(y,z)-1) +
    f^{-}_{\sigma}(y) \overline{f^{-}_{\sigma}}(z) ({\cal I}(y,z)-1)
    \right.\nonumber\\ && \left. +f^{+}_{\sigma}(y)
    \overline{f^{-}_{\sigma}}(z) ({\cal I}(y,z)-\Lambda) +
    f^{-}_{\sigma}(y) \overline{f^{+}_{\sigma}}(z) ({\cal I}(y,z)-
    \overline{\Lambda}) \right|^{2}\nonumber\\ &&\leq \sup_{y,z \in
    \Delta^{+}} |{\cal I}(y,z) -1|^{2} + \sup_{y,z \in \Delta^{-}}
    |{\cal I}(y,z) -1|^{2} \nonumber\\ &&+ \sup_{y \in \Delta^{+},z
    \in \Delta^{-}} |{\cal I}(y,z) -\Lambda |^{2} + \sup_{y \in
    \Delta^{-},z \in \Delta^{+}} |{\cal I}(y,z) -\overline{\Lambda}
    |^{2} \ea

    \n
    Using proposition 2 we conclude the proof.  $\Box$

    \vs
    \n
    From corollary 3 and theorem 1 we conclude that the reduced density
    matrix for the heavy particle in the position representation
     can be approximated by the density matrix

    \ba &&\hat{\rho}^{f}(t,R,R')= \f{1}{2} ( U_{0}^{M}(t)
    f^{+}_{\sigma})(R)  (U_{0}^{M}(-t) \overline{f^{+}_{\sigma}})(R')
    + \f{1}{2} (U_{0}^{M}(t) f^{-}_{\sigma})(R) ( U_{0}^{M}(-t)
    \overline{f^{-}_{\sigma}})(R') \nonumber\\ &&+\f{\Lambda}{2}
    (U_{0}^{M}(t) f^{+}_{\sigma})(R) ( U_{0}^{M}(-t)
    \overline{f^{-}_{\sigma}})(R') +\f{\overline{\Lambda}}{2}
    (U_{0}^{M}(t) f^{-}_{\sigma})(R) ( U_{0}^{M}(-t)
    \overline{f^{+}_{\sigma}})(R') \label{fin} \ea

    \n
    with an explicit control of the error.

    \n
    If the interaction with the light particle is switched off, i.e.
    for $\alpha =0$, we have $\Lambda =1$ and  then (\ref{fin})
    reduces to the pure state corresponding to the  coherent
    superposition of the free evolution of the two wave packets
    $f^{\pm}_{\sigma}$.

    \n
    On the other hand, if $\alpha >0$ one easily sees that $0<
    |\Lambda| <1$ and then  (\ref{fin}) is a mixed state for which the
    interference terms are   reduced by the factor $\Lambda$ and this
    is the typical manifestation of the (partial) decoherence effect
    induced by the light particle on the heavy one.

    \n
    The relevant parameter $\Lambda$ (see (\ref{Lambda})) is defined
    in terms of the probability distribution of the momentum of the
    light particle  $|\tilde{g}_{\delta}(k)|^{2}$ and of the
    transmission coefficient ${\cal T}_{\alpha}(k)$.

    \n
    Then the decoherence effect is emphasized if the fraction of
    transmitted wave for the light particle is small.


    \n
    In particular, rescaling the integration variable in
    (\ref{Lambda}), one can also write

    \be
    \Lambda = 1 - \int dz |\tilde{g}(z)|^{2} \frac{\alpha
    \delta}{\alpha \delta - i (z + k_{0} \delta)}\label{decoheff1} \ee

    \n
    Then one easily sees that for $k_0 \gg \alpha$ the light particle
    is completely transmitted and $\Lambda \simeq 1$, i.e. the
    decoherence effect is negligible.

    \n
    On the other hand, for $k_0 \ll \alpha$ the decoherence effect is
    nonzero and given by

    \be
    \Lambda \simeq \int dz |\tilde{g}(z)|^{2} |{\cal T}_{\alpha
    \delta}(z)|^{2} = 1 - \int dz |\tilde{g}(z)|^{2} \frac{\alpha^2
    \delta^2}{\alpha^2 \delta^2 + z^2}\label{decoheff} \ee

    \n
    where we have used the fact that $\tilde{g}$ is even.

    \n
    A further interesting question is the analysis of
    $\hat{\rho}^{f}(t)$ in the momentum representation.

    \n
    Since momentum is a constant of the motion, the density matrix is
    simply given by

    \ba &&\f{1}{\hbar}
    \tilde{\rho}^{f}_{0}\left(\f{P}{\hbar},\f{P'}{\hbar}\right)=
    \f{1}{2\hbar} \tilde{f}_{\sigma}^{+} \left(\f{P}{\hbar}\right)
    \overline{\tilde{f}_{\sigma}^{+}} \left(\f{P'}{\hbar}\right) +
    \f{1}{2\hbar} \tilde{f}_{\sigma}^{-} \left(\f{P}{\hbar}\right)
    \overline{\tilde{f}_{\sigma}^{-}} \left(\f{P'}{\hbar}\right)
    \nonumber\\ && + \f{\Lambda}{2\hbar} \tilde{f}_{\sigma}^{+}
    \left(\f{P}{\hbar}\right) \overline{\tilde{f}_{\sigma}^{-}}
    \left(\f{P'}{\hbar}\right) + \f{\overline{\Lambda}}{2\hbar}
    \tilde{f}_{\sigma}^{-} \left(\f{P}{\hbar}\right)
    \overline{\tilde{f}_{\sigma}^{+}} \left(\f{P'}{\hbar}\right)
    \label{fin-mom} \ea

    \n
    It is then clear that the decoherence effect is present also in
    the momentum representation and it is measured by the same
    parameter $\Lambda$.

    \n
    Moreover, if $\tilde{f}_{\sigma}^{+}$ and $\tilde{f}_{\sigma}^{-}$ are
    well separated,
    one easily realizes that the probability distribution of the momentum
    remains
    essentially unchanged with respect to the unperturbed case $\Lambda =1$,
    the error
    being of order $\epsilon$.

    \n
    We analyse now the evolution in the position representation of the
    heavy particle exploiting the approximate reduced density matrix
    $\hat{\rho}^{f}(t)$.

    \n
    We shall explicitely show that the typical interference fringes
    produced by the superposition state when the interaction with the
    light particle is  absent, i.e. for $\Lambda =1$, are in fact
    reduced when the light  particle is present, i.e. for
    $|\Lambda|<1$.

    \n
    In order to see the effect more clearly we assume

    \ba && \f{\sigma}{R_{0}} \ll \f{\hbar}{\sigma P_{0}} \label{spre}
    \ea

    \n
    The effect of the interfernce terms becomes more relevant when the
    supports
     of the two wave pakets $U_{0}^{M}(t)f^{\pm}_{\sigma}$ have the maximal
    overlapping and this approximately happens at the time $t=\tau \equiv
    \f{R_0 M}{P_0}$.
    Then, from (\ref{fin}), we consider

    \ba &&n(\tau,R) \equiv \hat{\rho}^{f}(\tau,R,R) \nonumber\\ &&=
    \f{1}{2} \left[ |(U_{0}^{M}(\tau) f_{\sigma}^{+})(R)|^{2} +
    |(U_{0}^{M}(\tau) f_{\sigma}^{-})(R)|^{2} + 2 \Re \left( \Lambda
    (U_{0}^{M}(\tau) f_{\sigma}^{+})(R) \overline{(U_{0}^{M}(\tau)
    f_{\sigma}^{-})}(R) \right)\right]\nonumber\\ && \label{probpos}
    \ea

    \n
    Using  (\ref{spre})  and a standard scattering estimate (see e.g.
    [RS]) we obtain

    \ba &&(U_{0}^{M}(\tau) f_{\sigma}^{+})(R)= \sqrt{\f{P_{0}
    \sigma}{2 \pi \hbar R_{0}}} \int dx f \left(\f{x + R_{0}}{\sigma}
    \right) e^{i \f{P_{0}}{\hbar}x + i \f{P_{0}}{2 \hbar R_{0}}
    (R-x)^{2}} \nonumber\\ &&= \sqrt{\f{P_{0} \sigma}{i \hbar R_{0}}}
    e^{i \f{P_{0}}{\hbar} \left( \f{R^{2}}{2 R_{0}} +R -\f{R_{0}}{2}
    \right) } \tilde{f}\left(\f{P_{0}\sigma}{\hbar R_{0}} R \right) +
    {\cal E}_0(R)\\ &&\|{\cal E}_0\| < \f{P_{0} \sigma^{2}}{2 \hbar
    R_{0}} \|\Delta \tilde{f}\| \ea

    \n
    Proceeding analogously for $(U_{0}^{M}(\tau) f_{\sigma}^{-})(R)$
    we find

    \ba &&n(\tau,R) = \f{P_{0}\sigma}{\hbar R_{0}} \left|
    \tilde{f}\left( \f{P_{0}\sigma}{\hbar R_{0}} R \right) \right|^{2}
    \left( 1 + |\Lambda| \cos \left( \f{2 P_{0}}{\hbar} R + \varphi
    \right) \right) + {\cal E}_{1}(R) \label{deco}\\ &&\|{\cal E}_1
    \|_{L^{1}}  < c \; \f{P_{0}\sigma^{2}}{\hbar R_{0}} \ea

    \n
    where

    \ba &&\Lambda = |\Lambda| e^{i \varphi} \ea

    \n
    For $|\Lambda| < 1$, formula (\ref{deco}) shows that the presence
    of the light particle determines a reduction of the amplitude of
    the oscillations
     and a shift of the corresponding phases.

    \n
    Notice that the shift is negligible if $k_{0} \ll \alpha$.

    \vspace{1cm}

    \setcounter{chapter}{4} \setcounter{equation}{0}

    \vs

    {\bf 4. Proof of theorem 1} \vs

    \vs

    \n
    The proof of theorem 1 will be obtained through the proof of three
    lemmas.

    \vs
    \n
    {\bf Lemma 4.} {\em Given the initial state (\ref{statoin}), for
    any $t \geq 0$ one has

    \be
    \|\psi(t)- \psi_1(t) \| < C_1 \,\epsilon \label{stima1} \ee

    \n
    where

    \be
    \psi_1(t,r,R)= \int  dy f_{\sigma} (y) U_0^{\nu} \left(t,
    \f{M}{\nu} R + \f{\mu}{M}r - y \right) \int dr' g_{\delta}(r' +y)
    U_{\alpha_{0}}^{\mu} \left(t, r-R, r' \right) \label{psi1} \ee

    \n
    and}

    \be
    C_1 = \left[ \int dx x^2 \int dy \left| \f{\partial}{\partial y} \left(
    f_{\sigma}(y) g_{\delta}(x+y)\right) \right|^{2} \right]^{1/2}
    \label{C_1}
    \ee


    \vs
    \n
    {\bf Proof}.  Using the relative and the center of mass
    coordinates (see (\ref{T})), from (\ref{solu.}) one has

    \be
    (T \psi(t))(x_1,x_2)= \left( U_0^{\nu}(t) U_{\alpha_{0}}^{\mu}(t)
    T \psi_0 \right)(x_1,x_2) \ee

    \n
    where $U_{\alpha_{0}}^{\mu}(t)$ is defined in (\ref{sol.delta}) of
    the appendix and

    \be
    (T \psi_0)(x_{1}, x_{2})= f_{\sigma}\left(x_{2} - \f{\mu}{M}
    x_{1}\right) g_{\delta} \left(x_{2} + \f{M}{\nu} x_{1} \right) \ee

    \n
    Moreover

    \ba &&(T\psi_{1}(t))(x_1,x_2) = \int dx_{2}' dx_{1}'
    f_{\sigma}(x_{2}') g_{\delta}(x_{1}' + x_{2}') U_{0}^{\nu}(t, x_2
    - x_{2}') U_{\alpha_{0}}^{\mu}(t,x_1,x_{1}')\nonumber\\ &&\equiv
    \left(   U_{0}^{\nu}(t)  U_{\alpha_{0}}^{\mu}(t) T \psi_{01}
    \right) (x_1,x_2) \ea

    \n
    where

    \be
    \psi_{01}(r,R)= f_{\sigma} \left( \f{M}{\nu} R + \f{\mu}{M} r
    \right) g_{\delta} \left(r - R + \f{M}{\nu} R + \f{\mu}{M}r
    \right) \ee

    \n
    Then we have with $\frac{\mu}{M}=\frac{\epsilon}{1+\epsilon}, \,
    \frac{M}{\nu}-1= -\frac{\epsilon}{1+\epsilon}$

    \ba &&\| \psi(t) - \psi_1(t)\|^{2} = \|T \psi (t) - T \psi_1(t)
    \|^{2}= \| T \psi_0 - T \psi_{01} \|^{2}\nonumber\\ &&=\int dx_1
    dx_2 \left|   f_{\sigma}\left(x_{2} - \f{\mu}{M} x_{1}\right)
    g_{\delta} \left(x_{2} + \f{M}{\nu} x_{1} \right) -
     f_{\sigma}(x_{2})
    g_{\delta}(x_{1} + x_{2}) \right|^{2}\nonumber\\ &&= \int dx_1
    dx_2 \left|   f_{\sigma}\left(x_{2} - \frac{\epsilon}{1+\epsilon}
    x_{1}\right) g_{\delta} \left(x_{2} + x_1 -
    \frac{\epsilon}{1+\epsilon} x_{1} \right) - f_{\sigma}(x_{2})
    g_{\delta}(x_{1} + x_{2}) \right|^{2} \nonumber\\ &&= \int dx_1
    dx_2 \left|F\left(x_1,x_2- \frac{\epsilon}{1+\epsilon}
    x_{1}\right) - F(x_1,x_2)\right|^2 \ea
    where $F(x_1,x_2)=
    f_{\sigma}(x_{2}) g_{\delta}(x_{1} + x_{2}).$ By a simple
    Plancherel argument, we have that

    \ba  && \int dx_1 dx_2 \left|F\left(x_1,x_2-
    \frac{\epsilon}{1+\epsilon} x_{1}\right) - F(x_1,x_2)\right|^2
    \nonumber\\
    &&= \int dx_1\int dk \left|\tilde{F}(x_1,k)\left(e^{-i
    \frac{\epsilon}{1+\epsilon} x_{1}k} - 1\right)\right|^2
    \nonumber\\
    &&\le   \int dx_1\int dk |\tilde{F}(x_1,k)|^{2}  \left(
    \frac{\epsilon}{1+\epsilon} x_{1}k\right)^2
    \nonumber\\
    &&=  \left(\frac{\epsilon}{1+\epsilon}\right)^2
    \int dx_1 x_1^2 \int dx_2 \left| \f{\partial}{\partial x_2}
    F(x_1,x_2)\right|^2
    \label{important}
    \ea
    from
    which the lemma follows. $\Box$

    \vs
    \n
    Since a small value of $m$ in the interacting unitary group
    $U_{\alpha_{0}}^{\mu}(t)$  is equivalent to a large value of $t$,
    in the next lemma we use a typical scattering estimate to
    approximate $U_{\alpha_{0}}^{\mu}(t)$ in (\ref{psi1}).

    \vs
    \n
    {\bf Lemma 5.} {\em Given the initial state (\ref{statoin}), for
    any $t>0$ one has

    \be
    \|\psi_1(t) - \psi_2(t) \| < \frac{C_2}{t} \, \epsilon \ee

    \n
    where

    \ba &&\psi_{2}(t,r,R)= \sqrt{\f{m}{2 \pi i \hbar t}} \sqrt{\f{M}{2
    \pi i \hbar t}} e^{i \f{m}{2 \hbar t} r^{2} + i\f{M}{2 \hbar
    t}R^{2}}
     \int d \xi f_{\sigma}(\xi)  e^{i \f{M}{2
    \hbar t} \xi^{2} } e^{-i \left( \f{M}{\hbar t}R + \f{m}{\hbar t}r
    \right) \xi}\nonumber\\ &&\cdot \int d r' g_{\delta}(r' + \xi)
    \left( e^{-i \f{\mu}{\hbar t} (r-R) r'} - \f{e^{i \f{\mu}{\hbar
    t}|r-R| |r'|}}{1- i \f{\hbar }{\alpha_{0} t} |r-R|}
    \right)\label{laststep} \ea

    \n
    and}

    \ba
    &&C_2 = c \f{M}{\hbar} \left\{ \int dx x^4 |f_{\sigma}(x)|^2  + \int dx
    |f_{\sigma}(x)|^2 \left[ \int dy y^4
    |g_{\delta}(y+x)|^2 + \f{1}{\alpha^3} \left( \int dy
    |g_{\delta}(y+x)|\right)^{2} \right. \right.\nonumber\\
    &&\left. \left. + \f{1}{\alpha} \left( \int dy |y|
    |g_{\delta}(y+x)|\right)^{2}+ \alpha \left(\int dy y^2 |g_{\delta}(y+x)|
    \right)^2
    \right] \right\}^{1/2}
    \label{C_2}
    \ea

    \vs
    \n
    {\bf Proof. } We shall first estimate the difference $\psi_1 (t) -
    \hat{\psi}_2 (t)$, where $\hat{\psi}_2 (t)$ is explicitely given by

    \ba &&\hat{\psi_2}(t,r,R)= \sqrt{\f{\mu}{2 \pi i \hbar t }} e^{i
    \f{\mu}{2 \hbar t} (r-R)^{2}} \int dy f_{\sigma}(y) U_0^{\nu}
    \left(t, \f{M}{\nu} R + \f{\mu}{M}r - y \right) \int dr'
    g_{\delta}\left(r' + y \right)\nonumber\\ &&\cdot \left( e^{-i
    \f{\mu}{\hbar t} (r-R) r'} - \f{e^{i \f{\mu}{\hbar t}|r-R|
    |r'|}}{1- i \f{\hbar }{\alpha_{0} t} |r-R|} \right) \label{psi2}
    \ea

    \n
    From (\ref{psi1}) we have

    \ba &&(T \psi_1)(t,x_1,x_2) = \int dy U_0^{\nu}(t,x_2 -y)
    \varphi_1(t,x_1,y)\\ &&\varphi_1(t,x_1,x_2)=f_{\sigma}(x_2) \int
    dr' g_{\delta}(r'+x_2) U_{\alpha_{0}}^{\mu}(t,x_1,r') \ea

    \n
    and analogously for $\hat{\psi}_2 (t)$ we write

    \ba &&(T\hat{\psi}_2(t))(x_1,x_2)= \int d y U_0^{\nu}(t,x_2 -y)
    \varphi_2(t,x_1,y) \\ &&\varphi_2(t,x_1,x_2)= \sqrt{\f{\mu}{2 \pi
    i \hbar t}} e^{i \f{\mu}{2 \hbar t} x_{1}^{2}} f_{\sigma}(x_2)
    \int dr' g_{\delta}(r'+x_2) \left( e^{-i \f{\mu}{\hbar t} x_1 r'}
    - \f{e^{i \f{\mu}{\hbar t}|x_1| |r'|}}{1- i \f{\hbar
    |x_1|}{\alpha_{0} t}} \right) \ea

    \n
    Using the isometric character of  the operators $T$ and
    $U_0^{\nu}(t)$ and the explicit expression of
    $U_{\alpha_{0}}^{\mu}(t)$ (see (\ref{sol.delta})), we have

    \ba &&\|\psi_1(t) - \hat{\psi}_2(t)\|^{2}=\|T \psi_1(t) - T \psi_2(t)
    \|^{2}= \|\varphi_1(t)- \varphi_2(t)\|^{2}\nonumber\\ &&\leq 2
    \int dx_1 dx_2 \left| f_{\sigma}(x_2) \int dr' g_{\delta}(r' +x_2)
    \left( U_{0}^{\mu}(t, x_1 - r') - \sqrt{\f{\mu}{2 \pi i \hbar t}}
    e^{i \f{\mu}{2 \hbar t} x_{1}^{2} - i \f{\mu}{\hbar t}x_1 r'}
    \right) \right|^{2} \nonumber\\ &&+ 2 \int dx_1 dx_2 \left|
    f_{\sigma}(x_2) \int dr' g_{\delta}(r' +x_2) \left( \f{\mu
    \alpha_{0}}{\hbar^{2}} \int_{0}^{\infty} du e^{- \f{\mu
    \alpha_{0}}{\hbar^{2}} u} U_{0}^{\mu}(t, u+|x_1|+|r'|)
    \right.\right.\nonumber\\ && - \left. \left. \sqrt{\f{\mu}{2 \pi i
    \hbar t}} \f{1}{1- i \f{\hbar |x_1|}{\alpha_{0} t}}
     e^{i \f{\mu}{2
    \hbar t} x_{1}^{2} +i \f{\mu}{\hbar t}|x_1| |r'|} \right)
    \right|^{2} \nonumber\\ &&\equiv (I) +  (II) \label{I+II} \ea

    \n
    A standard estimate for the free unitary group (see e.g.  [RS])
    gives

    \ba &&(I)= \f{\mu}{\pi \hbar t}\int  dx_2 |f_{\sigma}(x_2)|^{2}
    \int dx_1 \left| \int dr' g_{\delta}(r' +x_2) \left( e^{i
    \f{\mu}{2 \hbar t}r'^{2}} -1 \right) e^{-i \f{\mu}{\hbar t} x_1 r'
    } \right|^{2} \nonumber\\ &&= \f{1}{\pi} \int  dx_2
    |f_{\sigma}(x_2)|^{2} \int d \bar{x}_1 \left| \int dr'
    g_{\delta}(r' +x_2) \left( e^{i \f{\mu}{2 \hbar t}r'^{2}} -1
    \right) e^{-i \bar{x}_1 r' } \right|^{2} \nonumber\\ &&=2 \int
    dx_2 |f_{\sigma}(x_2)|^{2} \int dr' |g_{\delta}(r' + x_2)|^{2}
    \left| e^{i \f{\mu}{2 \hbar t} r'^{2} } -1 \right|^{2}\nonumber\\
    &&\leq \f{1}{2} \left(\f{\mu}{\hbar t} \right)^{2} \int  dx_2
    |f_{\sigma}(x_2)|^{2} \int dr' r'^{4} |g_{\delta}(r' +
    x_2)|^{2}\nonumber\\
    &&=  \f{\epsilon^{2} M^{2}}{2(1+
    \epsilon)^{2} \hbar^{2} t^{2}}   \int  dx_2
    |f_{\sigma}(x_2)|^{2} \int dr'
    r'^{4} |g_{\delta}(r' + x_2)|^{2}
    \label{I_1}
    \ea

    \n
    where in the second line of (\ref{I_1}) we used Plancherel
    theorem.

    \n
    Concerning $(II)$, we introduce the  change of variables

    \be
    v= \f{\mu \alpha_{0}}{\hbar^{2}} u, \;\;\;\; y_1= \f{\mu}{\hbar t}
    x_1 \ee

    \n
    and use the identity

    \be
    \int_{0}^{\infty} dv e^{-v +i \f{\hbar |x_1|}{\alpha_{0} t}v} =
    \f{1}{1-i \f{\hbar |x_1|}{\alpha_{0} t}} \ee

    \n
    Then

    \ba &&(II) =  \f{1}{\pi} \int  dx_2 |f_{\sigma}(x_2)|^{2} \int
    dy_1 \left| \int d r' g_{\delta} \left(r' + x_2  \right)  e^{i
    |y_1||r'|}
     \right. \nonumber\\
    &&\cdot \left. \left( e^{i \f{\mu}{2 \hbar t } r'^{2}}
    \int_{0}^{\infty} dv e^{-v +i \f{1+\epsilon}{\alpha}|y_1|v +i
    \f{(1+\epsilon)m }{2 \hbar t \alpha^{2}}v^{2} + i \f{m}{\hbar t
    \alpha}|r'|v} - \int_{0}^{\infty} dv e^{-v +i
    \f{1+\epsilon}{\alpha}|y_1|v} \right) \right|^{2}\nonumber\\
    &&\leq \f{2}{\pi} \int  dx_2 |f_{\sigma}(x_2)|^{2} \int dy_1
    \left| \int dr' g_{\delta} \left(r' + x_2 \right)  e^{i |y_1||r'|}
    e^{i \f{\mu}{2 \hbar t }r'^{2}}
       \int_{0}^{\infty} dv e^{-v +i
    \f{1+\epsilon}{\alpha}|y_1|v}\right.\nonumber\\ && \left. \cdot
    \left( e^{ i \f{(1+\epsilon)m}{2 \hbar t \alpha^{2}}v^{2} + i
    \f{m}{\hbar t \alpha}|r'|v} -1 \right) \right|^{2}\nonumber\\ &&+
    \f{2}{\pi} \int  dx_2 |f_{\sigma}(x_2)|^{2} \int dy_1 \left| \int
    d r' g_{\delta}\left(r' + x_2 \right)  e^{i |y_1||r'|} \left( e^{i
    \f{\mu}{2 \hbar t }r'^{2}} -1 \right) \f{1}{1-i
    \f{1+\epsilon}{\alpha}|y_1|}\right|^{2}\nonumber\\ &&\equiv (III)
    + (IV) \label{II} \ea

    \vs
    \n
    The estimate of $(IV)$ is trivial

    \ba &&(IV) \leq \f{1}{2 \pi} \left(\f{\mu}{\hbar t} \right)^{2}
    \int  dx_2 |f_{\sigma}(x_2)|^{2} \int dy_1 \f{1}{1+ \left(
    \f{1+\epsilon}{\alpha}\right)^{2} y_1^{2}} \left( \int d r' r'^{2}
    \left|g_{\delta}\left(r' + x_2 \right)
    \right|\right)^{2}\nonumber\\
    &&\leq \f{\epsilon^{2} M^{2}
    \alpha}{2 (1+ \epsilon)^{3} \hbar^{2} t^{2}} \int  dx_2
    |f_{\sigma}(x_2)|^{2}
     \left( \int dr' r'^{2}
    |g_{\delta}(r' + x_2)|
    \right)^{2}  \label{IV}
    \ea

    \n
    For the estimate of $(III)$ it is convenient to integrate by parts
    the integral in the variable $v$

    \ba
    &&(III)= \f{2}{\pi} \int  dx_2 |f_{\sigma}(x_2)|^{2} \int dy_1
    \f{1}{1+ \left( \f{1+\epsilon}{\alpha}\right)^{2} y_1^{2}} \left|
    \int dr' g_{\delta}(r' + x_2)  e^{i |y_1||r'|} e^{i \f{\mu}{2
    \hbar t }r'^{2}} \right.\nonumber\\
    &&\left. \cdot
    \int_{0}^{\infty} dv e^{-v +i \f{1+\epsilon}{\alpha}|y_1|v  + i
    \f{(1+\epsilon)m}{2 \hbar t \alpha^{2}}v^{2} + i \f{m}{\hbar t
    \alpha}|r'|v} \f{(1+\epsilon)\mu}{\hbar t \alpha} \left(\f{1+
    \epsilon}{\alpha}v + |r'| \right)\right|^{2}\nonumber\\
    &&\leq
    \f{2(1+\epsilon)\mu^{2}}{\pi \hbar^{2} t^{2} \alpha} \int  dx_2
    |f_{\sigma}(x_2)|^{2} \int dy_1 \f{1}{1+ y_1^{2}} \left[  \int d
    r' |g_{\delta}(r' + x_2 )| \int_{0}^{\infty}dv e^{-v} \left(\f{1+
    \epsilon}{\alpha}v + |r'|\right) \right]^{2} \nonumber\\
    &&\leq
    \f{2 \epsilon^{2} M^{2} }{(1+ \epsilon) \hbar^{2} t^{2} \alpha}
    \int dx_2 |f_{\sigma}(x_2)|^2 \left[ \int dr' |g_{\delta} (r' + x_2)|
    \left( \f{1+ \epsilon}{\alpha} + |r'| \right)
    \right]^{2}
    \label{III}
    \ea

    \n
    Finally, it remains to analise the difference $\psi_2 (t) - \hat{\psi}_2
    (t)$. Using the explicit expression of $U_{0}^{\nu}(t)$ we have

    \ba &&\psi_{2}(t,r,R) - \hat{\psi}_{2}(t,r,R)\nonumber\\
    &&= \sqrt{\f{m}{2 \pi i
    \hbar t}} \sqrt{\f{M}{2
    \pi i \hbar t}} e^{i \f{m}{2 \hbar t} r^{2} + i\f{M}{2 \hbar
    t}R^{2}}
     \int d \xi f_{\sigma}(\xi) \left( e^{i \f{M}{2
    \hbar t} \xi^{2} } - e^{i \f{\nu}{2
    \hbar t} \xi^{2} }\right) e^{-i \left( \f{M}{\hbar t}R + \f{m}{\hbar t}r
    \right) \xi}\nonumber\\ &&\cdot \int d r' g_{\delta}(r' + \xi)
    \left( e^{-i \f{\mu}{\hbar t} (r-R) r'} - \f{e^{i \f{\mu}{\hbar
    t}|r-R| |r'|}}{1- i \f{\hbar }{\alpha_{0} t} |r-R|}
    \right) \ea

    \n
    Exploiting Plancherel theorem and the fact that the operator (\ref{W}) is
    unitary one easily sees that

    \be
    \| \psi_2 (t) - \hat{\psi}_2 (t) \|^2 \leq \left( \f{M}{2 \hbar t}
    \right)^{2} \epsilon^{2} \int d \xi \xi^4 |f_{\sigma}(\xi)|^{2}
    \label{psi-hatpsi}
    \ee

    \n
     From
    (\ref{I+II}),(\ref{I_1}),(\ref{II}),(\ref{IV}),(\ref{III}),
    (\ref{psi-hatpsi}) we conclude the proof of the lemma. $\Box$

    \vs
    \n
    In the last step we approximate (\ref{psi2}) using the fact that
    the coordinates of the heavy particle are slowly varying with
    respect to the coordinates of the light one.

    \vs
    \n
    {\bf Lemma 6.} {\em Given the initial state (\ref{statoin}), for
    any $t > 0$ we have

    \be
    \| \psi_{2}(t) -\psi^{a}(t) \| <  C_{3}(t) \,\epsilon \ee

    \n
    where

    \ba
    && C_3(t) = \left[ \int dz z^2 \int dx \left| \f{\partial}{\partial x}
    \left(
    \tilde{\hat{f}}_{\sigma}(z-x) \tilde{g}(x) \right) \right|^{2}
    \right]^{1/2}\nonumber\\
    &&+ \f{1}{\sqrt{2} \pi} \left\{ \int dxdz \left[ \f{\alpha^{2} +
    z^{2}}{\alpha^{2}+x^{2}} \left( \int dr' |\zeta(z,r')|\right)^{2} +
    \f{\alpha^{2}z^{2}}{\alpha^{2} + x^{2}} \left(\int dr' |r'| |\zeta(z,r')|
    \right)^{2} \right] \right\}^{1/2}\label{C_3}
    \ea
  \n
  with

  \be
  C_3(t) < \f{C_4}{t} + C_5
  \label{C_4C_5}
  \ee

    \n
    and}

    \be
    \hat{f_{\sigma}}(\xi)=
    f_{\sigma}(\xi) e^{i \f{M}{2 \hbar t} \xi^{2} }, \;\;\;\;\;\;\;\;\;\;\;\;
    \zeta(z,r') = \int d\xi \hat{f}_{\sigma}(\xi) g_{\delta}(r' + \xi)  e^{-i
    z \xi}\label{hatf-zeta}
    \ee

    \vs

    \vs
    \n
    {\bf Proof.}  From (\ref{psiacompatta}) and (\ref{laststep}) we have

    \ba &&\psi_2 (t,r,R) - \psi^{a}(t,r,R)\nonumber\\
    &&= \sqrt{\f{m}{2 \pi i
    \hbar t}}
    \sqrt{\f{M}{2
    \pi i \hbar t}} e^{i \f{m}{2 \hbar t} r^{2} + i\f{M}{2 \hbar
    t}R^{2}} \int d \xi f_{\sigma}(\xi)  e^{i \f{M}{2 \hbar t} \xi^{2}
    } e^{-i \left( \f{M}{\hbar t}R + \f{m}{\hbar t}r \right)
    \xi}\nonumber\\ &&\cdot \left[ \int d r' g_{\delta}(r' + \xi)\left(e^{-i
    \f{\mu}{\hbar t}( r -R) r'}-  e^{-i
    \f{m}{\hbar t} r r'} \right) +
    \int d r' g_{\delta}(r' + \xi)\left( \f{e^{i\f{\mu}{\hbar t}|r-R| |r'|}}{1-
    i
    \f{\hbar }{\alpha_{0} t} |r-R|}-
    \f{e^{i\f{m}{\hbar t}|r| |r'|}}{1- i
    \f{\hbar }{\alpha_{0} t} |r|} \right) \right]\nonumber\\
    &&\equiv \left( \psi_2 -\psi^{a} \right)_{fr}(t,r,R) +
    \left( \psi_2 -\psi^{a} \right)_{in}(t,r,R)
    \label{psia} \ea

    \n
    We will estimate separatly the two terms $\left( \psi_2 -\psi^{a}
    \right)_{fr}(t)$ and $\left( \psi_2 -\psi^{a} \right)_{in}(t)$.

    \n
    Introducing the new integration variables

    \be
    x= \f{m}{\hbar t}r, \;\;\;\; z=\f{MR + mr}{\hbar t}
    \label{change-var}
    \ee

    \n
    and the function $\hat{f_{\sigma}}(\xi) $ defined in (\ref{hatf-zeta}), we
    have

    \ba &&\|  \left( \psi_2 -\psi^{a} \right)_{fr}(t) \|^{2} \nonumber\\
    &&= \f{1}{(2 \pi)^{2}} \int dx dz \left| \int d\xi
    \hat{f}_{\sigma}(\xi) e^{-iz \xi}
     \int d r' g_{\delta}\left(r' + \xi \right) \left(  e^{-i
    \left( x
    - \f{\epsilon}{1+\epsilon} z \right) r'} - e^{-i x r'}
    \right)\right|^{2}\nonumber\\
    &&=\f{1}{2 \pi} \int dx dz \left| \int
    d\xi
    \hat{f}_{\sigma}(\xi) e^{-iz \xi}\left(
    \tilde{g}_{\delta}\left( x
    - \f{\epsilon}{1+\epsilon} z\right)  e^{i \left( x
    - \f{\epsilon}{1+\epsilon} z\right) \xi } -
    \tilde{g}_\delta(x)e^{ix\xi}\right)\right|^{2}
    \nonumber\\
    && =   \int dx dz \left|
    \tilde{\hat{f_\sigma}} \left(z-x +\f{\epsilon}{1 +
    \epsilon} z \right)
    \tilde{g}_{\delta}\left(  x
    - \f{\epsilon}{1+\epsilon} z \right)
    - \tilde{\hat{f_\sigma}}(z-x)
    \tilde{g}(x)\right|^2\nonumber\\
    &&=\int dx dz \left|
    G \left(x-\f{\epsilon}{1+\epsilon}z, z \right)-G(x,z)\right|^2
    \ea

    \n
    where we introduced the function $  G(x,z)=\tilde{\hat{f_\sigma}}(z-x)
    \tilde{g}(x).$

    \n
    Proceeding as in (\ref{important}) we find

    \be
    \|  \left( \psi_2 -\psi^{a} \right)_{fr}(t) \|^{2} \leq
    \left(\f{\epsilon}{1+ \epsilon} \right)^{2} \int dz z^2 \int dx \left|
    \f{\partial}{\partial x} \left( \tilde{\hat{f_{\sigma}}}(z-x) \tilde{g}(x)
    \right) \right|^{2}
    \label{free}
    \ee



      \n
    For the estimate of $(\psi_2 - \psi^{a})_{in}(t)$ we use again  the
    change of variables (\ref{change-var})
      and we introduce the function $\zeta(z,r')$ defined in (\ref{hatf-zeta})

    \n
    Then we have

      \ba
      &&\|(\psi_2 - \psi^{a})_{in}\|^{2}\leq \f{1}{(2 \pi)^{2}} \int dx dz
    \left|
    \int dr'
    \zeta(z,r')
    \left(
    \frac{e^{i
      \left| x - \f{\epsilon}{1+ \epsilon}z\right| |r'|}}
      {1-\f{i}{\alpha} |(1+ \epsilon) x - \epsilon z|}
      - \f{e^{i|x||r'|}}
      {1 - \f{i}{\alpha} |x|}
      \right) \right|^{2}\nonumber\\
      &&\leq \f{1}{2 \pi^{2}} \int dx dz \left| \f{1}{1-\f{i}{\alpha} |(1+
    \epsilon) x -
    \epsilon
      z|}- \f{1}{1 - \f{i}{\alpha} |x|} \right|^{2}  \left( \int dr'
      |\zeta(z,r')|\right)^{2} \nonumber\\
      &&+  \f{1}{2 \pi^{2}} \int dx dz \f{\alpha^{2}}{\alpha^{2} + x^{2}}
    \left| \int dr'
      \zeta(z,r') \left( e^{i \left| x- \f{\epsilon}{1 +\epsilon}z\right||r'|}
    - e^{i|x||r'|} \right) \right|^{2} \nonumber\\
      &&\leq \f{\epsilon^{2}}{2 \pi^{2}}  \int dxdz \left[ \f{\alpha^{2} +
    z^{2}}{\alpha^{2} +
      x^{2}} \left( \int dr' |\zeta(z,r')| \right)^{2}  +
    \f{\alpha^{2} z^{2}}{\alpha^{2} +
      x^{2}}\left( \int dr' |r'| |\zeta(z,r')| \right)^{2}\right]
      \label{int1}
      \ea

      \n
      where we have used the estimates

      \ba && \left| \f{1}{1- \f{i}{\alpha} |(1+ \epsilon) x-\epsilon z
      |} - \f{1}{1- \f{i}{\alpha}|x|}\right|^{2} = \f{\alpha^{2}}{
      \alpha^{2} + x^{2}} \f{(|x|-|(1 + \epsilon)x - \epsilon
      z|)^{2}}{\alpha^{2} +  ((1 + \epsilon)x - \epsilon
      z)^{2}}\nonumber\\ &&\leq \epsilon^{2} \f{\alpha^{2}}{ \alpha^{2}
      + x^{2}} \f{(x - z)^{2}}{\alpha^{2} +  ((1 + \epsilon)x - \epsilon
      z)^{2}} \leq \left( \f{\epsilon}{1 + \epsilon} \right)^{2} \f{
      \alpha^{2} + z^{2}}{ \alpha^{2} + x^{2}} \label{omega} \ea

      \n
      and

      \be
      \left| e^{i \left| x - \f{\epsilon}{1+ \epsilon} z \right| |r'|} -
      e^{i|x||r'|} \right| \leq  \f{\epsilon}{1+ \epsilon} |z| |r'|
      \ee

      \n
      The function $\zeta(z,r')$ is smooth and, using repeated integration
      by parts, one easily sees that it is rapidly decreasing when its first
      argument goes to infinity. The computation is long but straightforward
      and we omit the details. The conclusion is that the integral in the last
    line of (\ref{int1}) is finite. Along the same line one can verify that
    (\ref{C_4C_5}) holds, where the constants $C_4$,
    $C_5$ are independent of time and then  the
  proof of the lemma follows. $\Box$

    \vs
    \n
    {\bf Proof of theorem 1.} It is an obvious consequence of lemmas
    4, 5, 6. $\Box$

    \vspace{1cm}

    \setcounter{chapter}{5} \setcounter{equation}{0}

    {\bf 5. Appendix: explicit solution of the two-body problem}

    \vs

    \n
    We recall here  the solution of the Schr\"{o}dinger equation

    \be
    i \hbar \f{\partial \psi(t)}{\partial t} = H \psi(t), \;\;\;\;
    \psi(0) = \psi_0 \label{Schr} \ee

    \n
    where $H$ is the self-adjoint hamiltonian in $ L^{2}(\R^{2},drdR)$
    given by (\ref{ham}). Using the unitary operator $T$ (see
    (\ref{T}) one  obviously has

    \be
    T H T^{-1} =
     H_{0}^{\nu} +  H_{\alpha_{0}}^{\mu}, \;\;\;\;  H_{0}^{\nu} =
     -\f{\hbar^{2}}{2 \nu} \Delta_{x_2},\;\;\;\;
    H_{\alpha_{0}}^{\mu} = - \f{\hbar^{2}}{2 \mu} \Delta_{x_1} +
    \alpha_{0} \delta(x_1) \ee

    \n
    where $(x_{1},x_{2})$ are the relative and the center of mass
    coordinates

    \be
    x_{1}= r-R,\;\;\;\; x_{2}= \f{mr + MR}{m+M} \ee

    \n
    Then the solution of (\ref{Schr}) can be written as

    \ba &&\psi(t,r,R) = \left( T^{-1} U_{0}^{\nu}(t)
    U_{\alpha_{0}}^{\mu}(t) T \psi_0 \right)(r,R)\nonumber\\ &&
    \nonumber\\ &&= \int dr' dR' \psi_{0}(r',R')
     U_{0}^{\nu} \left( t, \f{M}{\nu}(R-R') + \f{\mu}{M}(r-r') \right)
    U_{\alpha_{0}}^{\mu} \left(t, r-R,r'-R' \right) \nonumber\\ &&
    \label{solu.} \ea

    \n
    where the interacting unitary group $U_{\alpha_{0}}^{\mu}(t)$  is
    given by

    \be
    U_{\alpha_{0}}^{\mu}(t,x,x') = e^{-i \f{t}{\hbar}
    H_{\alpha_{0}}^{\mu}}(x, x'), \;\;\;\; x,x'\in \R \ee

    \n
    We remark that the evolution (\ref{solu.}) factorizes into a
    product of a free evolution in the center of mass coordinate and a
    one-body interacting evolution in the relative coordinate only if
    the initial state is of the form $\psi(r,R)=\psi_{1}(\f{M}{\nu} R+
    \f{\mu}{M}r) \psi_{2}(r-R)$.

    \vs

    \n
    In order to  compute $ (U_{\alpha_{0}}^{\mu}(t) \varphi_{0})(x)
    \equiv \varphi(t,x)$ one has to solve the one-body Schr\"{o}dinger
    equation

    \be
    i \hbar \f{\partial \varphi(t)}{\partial t}= - \f{\hbar^{2}}{2
    \mu} \Delta_{x} \varphi(t) + \alpha_{0} \delta(x)
    \varphi(t),\;\;\;\; \varphi(0)=\varphi_{0} \ee

    \n
    Defining the rescaled wave function

    \be
    \theta(s,z) =\varphi \left(\hbar s, \f{\hbar}{\sqrt{\mu}} z
    \right) \ee

    \n
    one finds that  $\theta(s)$ satisfies the corresponding equation
    with $\mu=\hbar=1$

    \be
    i  \f{\partial \theta (s)}{\partial s}= - \f{1}{2} \Delta_{z}
    \theta (s) + \alpha_{0}  \f{\sqrt{\mu}}{\hbar}  \delta(z) \theta
    (s),\;\;\;\; \theta (0)=\theta_{0} , \;\; \theta_{0}(z)=
    \varphi_{0}\left( \f{\hbar}{\sqrt{\mu}} z \right)
    \label{eq.Schul.} \ee

    \n
    The solution of (\ref{eq.Schul.}) can be found in [S]

    \be
    \theta(s,z)= \left( \hat{U}_{0}^{1}(s) \theta_{0} \right) (z) -
    \alpha_{0} \f{\sqrt{\mu}}{\hbar} \int_{0}^{\infty} dv e^{-
    \alpha_{0}  \f{\sqrt{\mu}}{\hbar} v} \int dz'
    \hat{U}_{0}^{1}(s,v+|z|+|z'|) \theta_{0}(z') \label{sol.Schul.}
    \ee

    \n
    where $\hat{U}_0^{1}(s)$ is the free propagator with $\hbar =1$.
    Noticing that $(U_{\alpha_{0}}^{\mu}(t) \varphi_{0})(x)=
    \varphi(t,x)=\theta(\f{t}{\hbar}, \f{\sqrt{\mu}}{\hbar}x)$ one has

    \ba && (U_{\alpha_{0}}^{\mu}(t) \varphi_{0})(x)=(U_{0}^{\mu}(t)
    \varphi_{0})(x) - \f{\mu \alpha_{0}}{\hbar^{2}} \int_{0}^{\infty}
    du e^{- \f{\mu \alpha_{0}}{\hbar^{2}} u} \int dx'
    U_{0}^{\mu}(t,u+|x|+|x'|) \varphi_{0}(x') \nonumber\\ &&
    \label{sol.delta} \ea

    \n
    Using (\ref{solu.}) and (\ref{sol.delta}), we finally obtain the
    complete solution (\ref{sol.esatta}) of the Schroedinger equation
    (\ref{Schr}).





    \vspace{2cm}

    {\bf References} 
\vs

    \n
    [ADGZ] Allori V., D\"urr D., Goldstein S., Zanghi, N.,   Seven Steps Towards 
the
  Classical World {\em Journal of Optics} {\bf B 4}, 482-488 (2002), 
quant-ph/0112005

    \n
    [AGH-KH] Albeverio S., Gesztesy F., Hoegh-Krohn R., Holden H.,
    {\em Solvable Models in Quantum Mechanics}, Springer-Verlag, 1988.

    \n
    [BGJKS]  Blanchard Ph., Giulini D., Joos E., Kiefer C., Stamatescu
    I.-O. eds.,{\em Decoherence: Theoretical, Experimental and
    Conceptual Problems}, Lect. Notes in Phys. 538, Springer, 2000.


    \n
    [DS] D\"urr D., Spohn H., Decoherence Through Coupling to the
    Radiation Field, in {\em Decoherence: Theoretical, Experimental
    and Conceptual Problems},  Blanchard Ph., Giulini D., Joos E.,
    Kiefer C., Stamatescu I.-O. eds., Lect. Notes in Phys. 538,
    Springer, 2000, pp. 77-86.


    \n
    [GF] Gallis M.R., Fleming G.N., Environmental and Spontaneous
    Localization, {\em Phys. Rev.} {\bf A42}, 38-48 (1990).

  \n
  [GH] Gell-Mann M., Hartle J. B., Classical equations for Quantum Systems,
  {\em Phys. Rev.} {\bf 47}, 3345-3382 (1993)

    \n
    [GJKKSZ] Giulini D., Joos E., Kiefer C., Kupsch J., Stamatescu
    I.-O., Zeh H.D., {\em Decoherence and the Appearance of a
    Classical World in Quantum Theory}, Springer, 1996.

    \n
    [H] Hagedorn G.A., A Time Dependent Born-Oppenheimer
    Approximation, {\em Comm. Math. Phys.} {\bf 77}, 1 (1980).

    \n
    [JZ] Joos E., Zeh H.D., The Emergence of Classical Properties
    Through Interaction with the Environment, {\em Z. Phys.} {\bf
    B59}, 223-243 (1985).


    \n
    [RS] Reed M., Simon B., {\em Methods of Modern Mathematical
    Physics, III: Scattering Theory}, Academic Press, 1979.

    \n
    [S] Schulman L.S., Application of the propagator for the delta
    function potential, in {\em Path Integrals from mev to Mev},
    Gutzwiller M.C., Ioumata A., Klauder J.K., Streit L. eds., World
    Scientific, 1986, pp. 302-311.

    \n
    [T] Tegmark M., Apparent Wave Function Collapse Caused by
    Scattering, {\em Found. Phys. Lett.} {\bf 6}, 571-590 (1993).

    \end{document}